\documentclass[twocolumn,prb,superscriptaddress,showpacs,floats]{revtex4}
\usepackage[dvips]{graphicx}
\usepackage{dcolumn}
\usepackage{bm}
\begin{document}

\title{\textit{Ab initio} study of the giant ferroelectric distortion and
pressure induced spin-state transition in BiCoO$_3$}

\author{Ting Jia}
\affiliation
{Key Laboratory of Materials Physics, Institute of Solid State Physics,
Chinese Academy of Sciences, Hefei 230031, China}
\author{Hua Wu}
\thanks{Corresponding author; wu@ph2.uni-koeln.de}
\affiliation
{II. Physikalisches Institut, Universit\"{a}t zu K\"{o}ln,
Z\"{u}lpicher Str. 77, 50937 K\"{o}ln, Germany}
\affiliation
{Department of Physics, Fudan University, Shanghai 200433, China}
\author{Guoren Zhang}
\affiliation{Key Laboratory of Materials Physics,
Institute of Solid State Physics, Chinese Academy of Sciences,
Hefei 230031, China}
\author{Xiaoli Zhang}
\affiliation{Key Laboratory of Materials Physics,
Institute of Solid State Physics, Chinese Academy of Sciences,
Hefei 230031, China}
\author{Ying Guo}
\affiliation{Key Laboratory of Materials Physics, Institute of Solid
State Physics, Chinese Academy of Sciences, Hefei 230031, China}
\author{Zhi Zeng}
\thanks{Corresponding author; zzeng@theory.issp.ac.cn}
\affiliation{Key Laboratory of Materials Physics,
Institute of Solid State Physics, Chinese Academy of Sciences,
Hefei 230031, China}
\author{Hai-Qing Lin}
\affiliation{Department of Physics and Institute of
Theoretical Physics, The Chinese University of Hong Kong, China}

\date{\today}

\begin{abstract}

Using configuration-state-constrained electronic structure calculations 
based on the generalized gradient approximation plus Hubbard $U$ method,
we sought the origin of the giant tetragonal ferroelectric distortion in
the ambient phase of the potentially multiferroic material BiCoO$_3$ 
and identified the nature of the pressure induced spin-state transition.
Our results show that a strong Bi-O covalency drives the giant ferroelectric
distortion, which is further stabilized by an $xy$-type orbital ordering of
the high-spin (HS) Co$^{3+}$ ions. For the orthorhombic phase under 5.8 GPa, 
we find that a mixed HS and low-spin (LS) state is more stable than both LS 
and intermediate-spin (IS) states, and that the former well accounts for the 
available experimental results. Thus, we identify that the pressure induced 
spin-state transition is via a mixed HS+LS state, and we predict that the 
HS-to-LS transition would be complete upon a large volume
decrease of about 20\%.  

\end{abstract}

\pacs{75.30.-m, 71.20.-b, 71.27.+a, 71.15.Mb}

\maketitle

\section{Introduction}

Multiferroic materials, having coexisting magnetism and ferroelectricity,
are of great technological and fundamental 
importance,~\cite{Cheong,Ramesh,Khomskii1} given the prospect
of controlling charges by applying magnetic fields and spins by voltages.
BiCoO$_3$ was recently synthesized by a high-pressure (HP) 
technique,~\cite{Belik} and it has been suggested to be a promising multiferroic
material by Uratani $et$ $al.$~\cite{Uratani} and by Ravindran 
$et$ $al.$~\cite{Ravindran} both through first-principles Berry-phase calculations. 
BiCoO$_3$ has a giant tetragonal
lattice distortion of $c$/$a$ = 1.27 with remarkable off-center atomic
displacements (see the inset of Fig. 1), and it is an insulator having C-type 
antiferromagnetism below
470 K---the antiferromagnetic (AF) $ab$ layers stacking ferromagnetically along 
the $c$ axis.~\cite{Belik}

It was proposed~\cite{Uratani,Oka,Okuno} that the giant tetragonal
distortion originates from lifting of the orbital degeneracy of the high-spin
(HS, $S$=2) Co$^{3+}$ ions and is stabilized by the subsequent $xy$-type
ferro-orbital ordering. Note that orbitally degenerate transition-metal
oxides quite often display an orbital ordering (OO) but ferroelectric (FE) materials
out of them are rare, as ferroelectricity and magnetism seem, 
and actually in most cases, 
to exclude each other.~\cite{Hill,Khomskii2}
Therefore, the proposed mechanism for the giant FE distortion appears
not straightforward.
Using fixed-spin-moment density-functional calculations,
Ravindran $et$ $al.$~\cite{Ravindran} predicted that there is a giant magnetoelectric
coupling in BiCoO$_3$: an external electric field (or a small
volume compression of $\sim$5\%) can induce a strong magnetic
response by changing the magnetic Co$^{3+}$-HS state in the 
FE phase into a nonmagnetic
low-spin (LS, $S$=0) state in a paraelectric (PE) phase.
A corresponding HS-insulator/LS-metal
transition was also suggested.~\cite{Ravindran,Ming} In sharp contrast, a very recent
HP study~\cite{Oka} showed 
that BiCoO$_3$ even under 6 GPa with a large volume decrease of 18\% is still semiconducting.
Note however that controversial spin states, both LS and intermediate-spin (IS, $S$=1),
were suggested for the HP phase.~\cite{Oka}

In the present work, we seek the origin of the giant tetragonal FE 
distortion in the ambient phase
of BiCoO$_3$ and identify the nature of the pressure induced
spin-state transition, using two sets of configuration-state-constrained 
GGA+$U$ (generalized gradient approximation plus Hubbard $U$) calculations.
Our results show that the giant tetragonal distortion is driven by 
a strong Bi-O covalency (rather than by the aforementioned
lifting of orbital degeneracy) and is further stabilized by an $xy$-type OO of the 
HS Co$^{3+}$ ions.
Moreover, we find that the pressure-induced spin-state transition is via a
mixed HS+LS state, which accounts for the available experimental
results consistently and disproves a recent prediction of a readily switchable
HS-LS transition.

\section{Computational Details}

We used the structural data of BiCoO$_{3}$ measured by the neutron
power diffraction.~\cite{Oka} Our calculations were performed 
using the full-potential augmented plane-wave plus local-orbital
code WIEN2k.~\cite{Blaha} The muffin-tin sphere radii were chosen to
be 2.3, 1.9, and 1.4 bohr for Bi, Co, and O atoms, respectively
(1.0 bohr for O when calculating the $E$-$V$ curves shown in Fig. 6).
The cutoff energy of 16 Ryd was set for the plane-wave expansion of 
interstitial wave functions, and 1200 {\bf k} points in the first 
Brillouin zone for the ambient structure with one formula unit (f.u.) and 
300 {\bf k} points for the HP phase with 4 f.u.

\begin{figure}
  \includegraphics[angle=0,width=6cm]{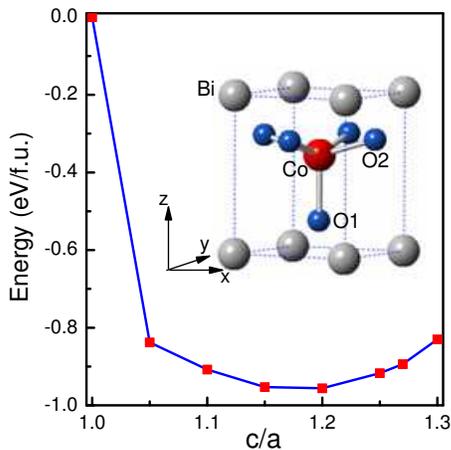}
  \caption{(Color online) $E$ $vs$ $c$/$a$ curve calculated by GGA+$U$ for the Co$^{3+}$-LS 
relaxed structures of BiCoO$_3$ with $c$/$a$ = 1-1.3 (in a step of 0.05)
including 1.27 (expt.). The inset shows a unit cell of the experimental
tetragonal structure.} 
  \label{fgr:1}
\end{figure}

Plain GGA [or local-spin-density approximation (LSDA)]
calculations~\cite{Uratani,Ravindran,Ming,Cai} seem to qualitatively
reproduce the C-type AF and insulating ground state of BiCoO$_3$ in the
ambient phase,
which was ascribed to the strong Hund-exchange stabilized HS state
of the Co$^{3+}$ ions (and thus the AF order and narrow bands) and
to the well split-off $xy$-singlet orbital.
However, the band gap of 0.6 eV and the Co$^{3+}$ spin moment of 2.4 $\mu_B$
given by the GGA/LSDA calculations are much smaller than the experimental
values of 1.7 eV and 3.2 $\mu_B$.\cite{Belik,McLeod}
Moreover, a recent prediction of an insulator-metal transition in 
BiCoO$_3$ upon a volume decrease of $\sim$5\% made by GGA/LSDA 
calculations~\cite{Ravindran,Ming} has already been disproved by
a very recent HP study,~\cite{Oka} which shows that BiCoO$_3$ is
still semiconducting even under 6 GPa with a large volume decrease of 18\%. 
As seen below, the experimental values of both the band gap and the Co$^{3+}$ spin
moment are well reproduced by our GGA+$U$
calculations. Note also that BiCoO$_3$ has an apparent AF order 
up to 470 K.\cite{Belik}
All these suggest that BiCoO$_3$ should rather be categorized as a
Mott insulator, with its band gap determined primarily by Hubbard $U$, i.e.,
strong correlation of the Co $3d$ electrons. 

To account for the strong electronic correlation,~\cite{Anisimov} we have carried out
GGA+$U$ calculations throughout this paper. 
In particular, we used the configuration-state-constrained GGA+$U$
method,~\cite{Korotin,Knizek,Wu1} which allows us to access different 
spin and orbital configuration states
of the concern by initializing their corresponding density matrix
and then doing self-consistent electronic relaxation.~\cite{Wu2}
This method is quite useful for study of the spin and orbital
physics present in transition-metal oxides.~\cite{Korotin,Knizek,Wu1,Wu2} All the
results shown below are obtained with $U$=6 eV and Hund exchange
$J$=0.9 eV. We note that our test calculations using other
reasonable $U$ values,~\cite{Cai,Wu1,Wu2} $U$=5 and 7 eV, gave
qualitatively the same results.

\section{Results and Discussion}

We first seek the origin of the giant tetragonal distortion 
by starting with our calculations assuming an ideal cubic structure
(space group $Pm\bar{3}m$) having the same volume as the experimental ambient
tetragonal structure. Then we changed the $c$/$a$ ratio of the tetragonal structure 
(keeping the volume unchanged) in our calculations. All those calculations
were carried out by setting a hypothetical LS state of the Co$^{3+}$ ions, and
by doing a full electronic and atomic relaxation for the cases of 
$c$/$a$ = 1.05-1.3 in a step of 0.05, and for the experimental 1.27 as well. 
All the solutions are insulating.
As the LS Co$^{3+}$ has a closed sub-shell $t_{2g}^6$
(thus no orbital degeneracy) and is an isotropic ion, we can use this
set of LS-constrained GGA+$U$ calculations as a computer experiment to probe mainly
the Bi-O covalent effect.

\begin{table}
  \caption{The total energies of BiCoO$_3$
(in unit of eV/f.u.) relative to the hypothetical cubic-structure LS state
calculated by GGA+$U$:
the experimental tetragonal phase assuming an LS state with atomic relaxations
(the third row),
assuming an HS state in the LS relaxed structure (the fourth row), and assuming the HS state
with further atomic relaxations (the fifth (last) row). The optimized atomic $z$-coordinates are also shown. 
Note that the structural data
of the atomically relaxed HS state are in good agreement with the
experiment~\cite{Oka} (see the inset of Fig. 1). 
See more discussion in the main text.}
  \label{tbl:1}
\begin{tabular}{lcc} \hline
States & $\Delta$$E$ & $z_{\rm Co}$, $z_{\rm O1}$, $z_{\rm O2}$ \\ \hline
Cubic LS & 0 & \\
Tetragonal LS-relaxed & --0.90 & 0.5897, 0.1889, 0.6853 \\
Tetragonal HS at LS-relaxed & --1.47 & \\
Tetragonal HS-relaxed & --1.97 & 0.5631, 0.1912, 0.7237 \\
 \hline
\end{tabular}
\end{table}

We show in Fig. 1 the calculated total energies as function
of the $c$/$a$ ratios, and one can immediately find that the $c$/$a$ ratio in the
LS equilibrium state is close to 1.2 (about 1.18), already indicating a large 
tetragonal distortion. The corresponding energy gain, relative to the hypothetical 
cubic structure ($c$/$a$ = 1), is 0.96 eV/f.u. For the LS relaxed structure with the 
experimental $c$/$a$ = 1.27, the energy gain is 0.90 eV/f.u.
We list in Table I (see the third row) the optimized atomic $z$-coordinates of the 
LS state with the experimental $c$/$a$ ratio,
and we find that the experimental CoO$_5$ coordination is well achieved 
[as seen from the optimized Co-O bondlengths, Co-O1: 1.893 \AA$\times$1, 
2.832 \AA$\times$1 (much larger); Co-O2: 1.919 \AA$\times$4] 
even in the presence of the Co$^{3+}$ LS state without orbital degeneracy.
This is also the case for $c$/$a$ = 1.2 with another optimized Co-O bondlengths,
1.881 \AA$\times$1, 2.677 \AA$\times$1, and 1.943 \AA$\times$4.
We plot in Fig. 2(a) a charge density contour of the hypothetical cubic structure
in the (100) plane, and in Fig. 2(b) that of the LS relaxed state with the 
experimental $c$/$a$ ratio. Fig. 2(a) shows nearly spherical charge densities 
around both the Bi and O ions, indicative of Bi$^{3+}$-O$^{2-}$ ionic bonds 
in the hypothetical cubic structure. In contrast, Fig. 2(b) clearly shows directional
Bi-O covalent bonds, which are also indicated the orbitally resolved density of 
states (see Fig. 3 and the discussion below). 
All these suggest that the Bi-O covalency drives the giant tetragonal 
FE distortion 
and causes remarkable off-center atomic displacements, no matter that 
the Co$^{3+}$ ions are in
this hypothetical LS state or in the real HS state (see below).
This is similar to the stereochemical mechanism of the 
Bi$^{3+}$ 6s$^2$ lone-pairs proposed for the highly distorted perovskite 
manganite BiMnO$_3$.~\cite{Seshadri} 

\begin{figure}[h]
 \includegraphics[angle=270,width=7cm]{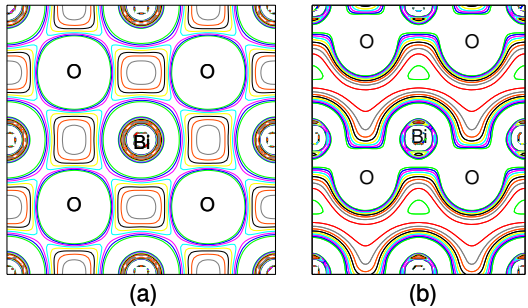}
  \caption{(Color online) Charge density contour (0.01-0.1 $e$/\AA$^3$ in a step of 0.01 $e$/\AA$^3$) 
of BiCoO$_3$ in the 
(100) plane of (a) the hypothetical ideal cubic structure and of (b) the 
LS-relaxed structure with the experimental $c$/$a$ ratio. 
The directional Bi-O covalency is apparent in (b).}
  \label{fgr:2}
\end{figure}

\begin{figure}[h]
 \includegraphics[angle=0,width=5.5cm]{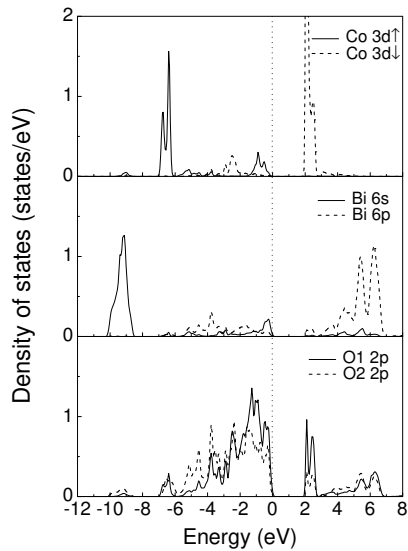}
  \caption{Partial density of states of BiCoO$_3$ in the C-type AF ground state of the
ambient phase.}
  \label{fgr:3}
\end{figure}

Furthermore, we study the effect of the $xy$-type OO of the HS Co$^{3+}$ ions
on the giant tetragonal distortion. 
Owing to the effective
CoO$_5$ pyramidal coordination already present in the LS relaxed tetragonal structures
as described above,
the crystal field (particularly the $t_{2g}$-$e_g$ splitting) is weak and thus 
less important than the Hund
exchange.~\cite{Hu} As a result, an HS state with an OO of the well 
split-off $xy$ orbital
is the ground state and more stable than the hypothetical LS state by 0.57 eV/f.u. for the
experimental $c$/$a$ = 1.27 (see the third and fourth rows in Table I) and by 0.59 eV/f.u. 
for $c$/$a$ = 1.2.
In this sense, the $xy$ OO is a consequence of the Bi-O covalency
driven tetragonal distortion and the associated CoO$_5$ pyramidal coordination.
Moreover, our calculations doing structural optimization show that an adjustment of 
the lattice to this $xy$ OO of the HS Co$^{3+}$ ions
changes the $c$/$a$ ratio to 1.28 (as compared to 1.18 for the LS equilibrium state, 
see Fig. 1). This agrees very well with the experimental $c$/$a$ = 1.27. 
Correspondingly, the experimental atomic parameters are also well reproduced, see the last row
in Table I. Furthermore, we show in Fig. 3 the density of states of BiCoO$_3$ in the 
C-type AF ground state having the $xy$ OO of the HS Co$^{3+}$ ions. The calculated band gap
of 1.98 eV and the Co$^{3+}$ spin moment of 3.01 $\mu_B$ are also in good agreement with
the experimental values of 1.7 eV and 3.24 $\mu_B$.\cite{Belik,McLeod} 
Hybridizations between the Bi $6s6p$
and O $2p$ orbitals are also apparent and evidence again the Bi-O covalency, 
although the magnitude of their respective
density of states is much underestimated within the muffin-tin spheres as those orbitals
are spatially quite spread. In a word, all above results allow us to conclude that 
the giant 
tetragonal FE distortion of BiCoO$_3$ originates from the Bi-O covalency 
(rather than from lifting of the orbital degeneracy of the HS Co$^{3+}$ 
ions~\cite{Uratani,Oka,Okuno}) and is further stabilized by the subsequent $xy$-type OO.

We now identify the nature of the pressure induced spin-state
transition in BiCoO$_3$, using another set of constrained
GGA+$U$ calculations for the orthorhombic structure measured at 5.8 GPa, by assuming
the LS, IS, and the mixed HS+LS states, respectively, and by doing a full electronic 
and atomic relaxation for each case.

\begin{table}
  \caption{The relative total energies $\Delta$$E$ (meV/f.u.), Co$^{3+}$ spin
moments $M$ ($\mu$$_{B}$), and band gap \emph{E$_{g}$} (eV) of
BiCoO$_{3}$ in the 5.8 GPa phase calculated by GGA+$U$ for
the LS-relaxed, IS-relaxed, and (HS+LS)-relaxed structures.}
  \label{tbl:2}
\begin{tabular}{lccc}
 \hline
States&$\Delta$$E$&$M_{\rm Co}$&\emph{E$_{g}$}\\
 \hline
(HS+LS)-relaxed&0&3.07, 0.17&0.84\\
LS-relaxed&91&0&0.95\\
IS-relaxed&158&2.01&half-metal\\
 \hline
\end{tabular}
\end{table}

\begin{figure}[h]
 \includegraphics[angle=0,width=6cm]{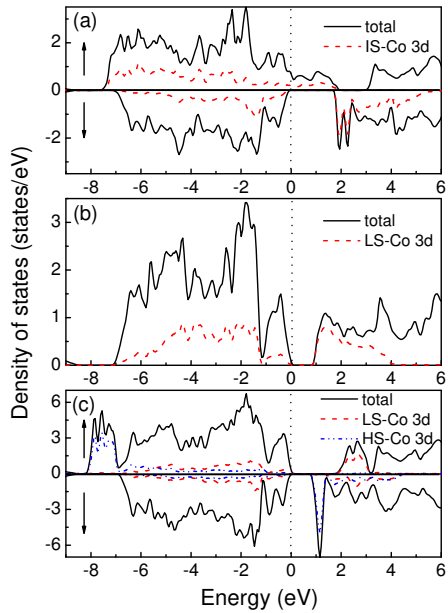}
  \caption{(Color online) The total and Co $3d$ density of states of BiCoO$_3$ in the 5.8 GPa phase
calculated by GGA+$U$ for the relaxed IS (a), LS (b), and mixed HS+LS (c) structures, 
respectively.}
  \label{fgr:4}
\end{figure}

As seen in Table II, the mixed HS+LS state has the lowest total energy, and the LS (IS)
state lies above it by 91 (158) meV/f.u. Those results suggest that
either the pure LS or IS state present in the 5.8 GPa phase~\cite{Oka} is not the case. 
Reversely, if the pure LS state were present in the 5.8 GPa phase, 
it would give rise to a change of the spin state from the pure HS state in
the ambient phase, $\Delta$$S$=2, being in disagreement with the
observed $\Delta$$S$=1.\cite{Oka} Moreover, an absence of the IS state is also not 
surprising, as (1) BiCoO$_3$ in
the 5.8 GPa phase is free of a Jahn-Teller distortion (that is expected for the localized IS
Co$^{3+}$); 
(2) a half-metallic band structure of the IS state (see Fig. 4(a))
disagrees with the measured semiconducting
behavior;~\cite{Oka} and (3) up to now a definite example of the insulating IS state appears 
still lacking, and even in the layered perovskites LaSrCoO$_4$ and 
La$_{1.5}$Sr$_{0.5}$CoO$_4$ both
having a strong tetragonal elongation of the Co$^{3+}$O$_6$ octahedra, the IS state 
turns out not to be the ground state either,~\cite{Wu2} despite an IS state 
might be intuitively expected.

As such, the mixed HS+LS state is most probably present in the 5.8 GPa phase.
An ideal 1:1 configuration of the mixed HS+LS state, with an average $S$=1, well
accounts for the observed change of the spin state.~\cite{Oka}
Moreover, a G-type order of the HS and LS Co$^{3+}$ ions (each HS Co
ion is surrounded by six LS Co ions, and vice versa), due to a
bigger/smaller size of the HS/LS Co$^{3+}$ ions, could help to gain
an elastic energy, and the resultant Co-O bondlengths are calculated to be
1.980 (1.930), 1.976 (1.890), and 1.948 (1.906) \AA~ for the HS (LS)
Co$^{3+}$ ions along the local $xyz$ axes.
Note, however, that a long-range G-type
order of the HS and LS Co$^{3+}$ ions is hard to establish, as only
single transition-metal species in a single charge state and in the
identical octahedral coordinations, i.e., solely Co$^{3+}$ ions are involved. 
Thus, the average Co-O bondlengths of the mixed and disordered HS+LS state 
(possibly with a short-range order due to a partial release of the lattice elasticity)
are also in good agreement with
the experiment.~\cite{Oka} Furthermore, the calculated insulating gap of
BiCoO$_3$ is reduced from 1.98 eV in the C-type AF state of the
ambient phase to 0.84 eV in the mixed HS+LS state (see Fig. 4(c)), which
qualitatively accounts for the decreasing resistivity of BiCoO$_3$
under pressure.~\cite{Oka}

\begin{figure}
 \includegraphics[angle=0,width=6cm]{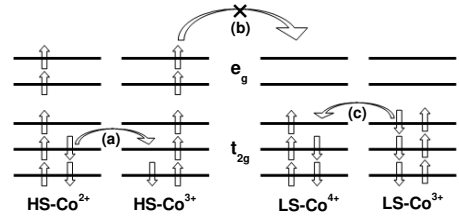}
  \caption{Electron hopping (a) from an HS Co$^{2+}$ ion
           to a neighboring HS Co$^{3+}$ and (c) from an LS Co$^{3+}$ to
           an LS Co$^{4+}$, but a suppressed electron hopping (b) 
           from an HS Co$^{3+}$ to an LS Co$^{4+}$ due to a spin blockade.} 
  \label{fgr:5}
\end{figure}

Note that when considering thermal
excitation of electrons into the initially empty conduction band and
holes left in the valence band for a nominally stoichiometric
material, these electron excitations would in a localized picture
correspond to HS Co$^{2+}$ states and holes to LS Co$^{4+}$. 
They behave like HS Co$^{2+}$ and LS Co$^{4+}$ ``impurities" in the matrix
of the Co$^{3+}$ ions. 
As a result, in the ambient phase of BiCoO$_3$ having the HS Co$^{3+}$ matrix, 
only the HS Co$^{2+}$ ``impurities" can transfer their
minority-spin $t_{2g}$ electrons to the neighboring HS Co$^{3+}$,
without changing the configuration states (the initial and
final states are the same), see Fig. 5(a). However, a hole hopping from
the LS Co$^{4+}$ to the HS Co$^{3+}$ is significantly
suppressed (see Fig. 5(b)), due to a cost of the Hund exchange energy associated
with a large change of the spin states which is referred as to a
spin-blockade machanism.~\cite{Maignan} In contrast, in the HP phase having
the mixed and disordered HS+LS Co$^{3+}$ matrix,
a charge hopping can take place both
between the HS Co$^{2+}$ and HS Co$^{3+}$ (Fig. 5(a)) and between the LS
Co$^{4+}$ and LS Co$^{3+}$ (Fig. 5(c)). This could also account for the
decreasing resistivity of BiCoO$_3$ under pressure.

Our above results show that
even in the 5.8 GPa phase of BiCoO$_3$ with a large volume decrease of 18\%,
the HS-to-LS
transition is not yet complete, and the system is most probably in the mixed HS+LS
insulating state, but not in the pure-LS metallic state which was predicted by
the previous GGA/LSDA calculations for BiCoO$_3$ upon a volume decrease of
$\ge$5\%.~\cite{Ravindran,Ming} 
As we discussed above, BiCoO$_3$ is rather a Mott insulator,
and thus the previous GGA/LSDA results are somewhat questionable: 
particularly the predicted
metallic solution was already disproved by a very recent experiment.~\cite{Oka}
Recent fixed-spin-moment calculations even predicted that BiCoO$_3$ could have a giant
magnetoelectric coupling with a readily switchable HS-LS transition
associated with an electric field driven FE-PE transition.~\cite{Ravindran}
We note that
the prediction may simply be an artifact of the fixed-spin-moment calculations,
as (1) LSDA or GGA (it was mentioned as a density-functional method in 
Ref~\cite{Ravindran})
is not well suitable to describe this
Mott insulator; (2) most probably those calculations were carried out in a wrong
ferromagnetic metallic state;
and (3) their metallic solutions blurred the distinction between
the different spin and orbital multiplets of the concern and thus suppressed
significantly their level
splittings and particularly the HS/LS splitting: the fixed-spin-moment
calculations showed that
for the ambient structure, the energy preference of the HS state over the LS state is 
less than 0.15 eV/f.u. (see Fig. 2 in Ref~\cite{Ravindran}),
whereas the corresponding value we calculated is more than 0.5 eV/f.u. Therefore, the previous
prediction of the HS-to-LS transition in BiCoO$_3$ with a small volume decrease of $\sim$5\%
was overly optimistic.~\cite{Ravindran} Instead, we find now that there is no readily switchable HS-LS
transition is BiCoO$_3$.

\begin{figure}
 \includegraphics[angle=0,width=6cm]{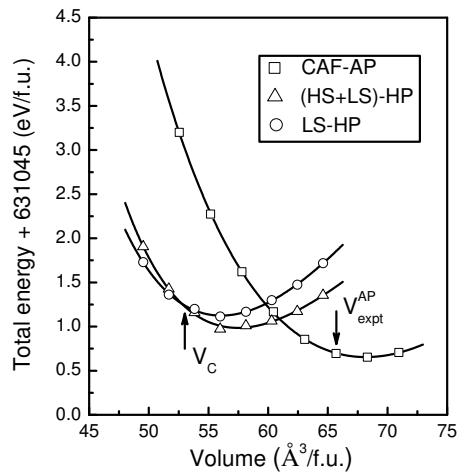}
  \caption{$E$-$V$ curves calculated by GGA+$U$ for the C-type AF state of BiCoO$_3$ 
in the ambient phase, and for the mixed HS+LS state and the LS state under high pressures. 
The lattice volume of the ambient phase is slightly overestimated within 3\%, and
the critical volume for a complete transition into the LS state is estimated to be
about 53 \AA$^{3}$/f.u.}
  \label{fgr:6}
\end{figure}

As Bi$^{3+}$ has a very similar ionic size as La$^{3+}$, 
it is reasonable to assume that when the local Co-O bondlengths of BiCoO$_3$
in the HP PE phase become identical to those of LaCoO$_3$ in the LS state at low 
temperature (1.925 \AA~ at 5 K),~\cite{Radaelli}
a complete transition into the LS state would be achieved in BiCoO$_3$. 
This, together with the structural data of the 5.8 GPa phase,\cite{Oka}
allows us to estimate the critical volume to be about 52.8 \AA$^{3}$/f.u. 
(corresponding to a volume decrease of about 20\%). 
Then, by extrapolating the eye-guided line of the $V$-$P$ data points
in the range of 2-6 GPa (see Fig. 1(b) in Ref~\cite{Oka}), we may estimate the 
critical pressure to be about 8 GPa.
It is important to note that the estimated critical volume is indeed well reproduced 
by our detailed calculations of the $E$-$V$ curves 
(see Fig. 6, $V_{\rm c}\approx$ 53 \AA$^{3}$/f.u.), 
which also nicely reproduce the equilibrium volume of the ambient phase within 3\%
and clearly indicate the HS-to-LS transition via the mixed HS+LS state.
This prediction of a complete transition into the LS state
awaits a further HP study.

\section{Conclusion}

To conclude, using configuration-state-constrained GGA+$U$ calculations,
we demonstrate that the giant tetragonal ferroelectric distortion of BiCoO$_3$ is driven 
by the strong Bi-O covalency (rather than by lifting of the 
orbital degeneracy of the HS Co$^{3+}$ ions) and is further stabilized by 
a subsequent $xy$-type OO. 
Moreover, our results show that the pressure induced HS-to-LS transition 
is via a mixed HS+LS state, 
and that the transition would be complete upon a large volume decrease of 
about 20\% (under about 8 GPa). The mixed HS+LS state well accounts for the 
available experimental results.\cite{Oka}

\section{Acknowledgments}

H. Wu is supported by the DFG via SFB 608 and by Fudan Univ.
The research at Hefei is funded by the NSF of China (Grant No. 11004195), 
the Special Funds for Major State Basic Research Project of China (973) 
under Grant No. 2007CB925004, Knowledge Innovation Program of CAS
(Grant No. KJCX2-YW-W07), Director Grants of CASHIPS, and CUHK
(Grant No. 3110023).

\end{document}